\documentclass[a4paper,11pt]{article}
\usepackage{pos}
\usepackage{amsmath}
\begin{document}
\title{Stochastic normalizing flows for lattice field theory}
%% \ShortTitle{Short Title for header}

\author[a,b]{M. Caselle}
\author*[a,b]{E. Cellini}
\author[a]{A. Nada}
\author[a,b]{M. Panero}

\affiliation[a]{Department of Physics,  University of Turin,\\
  Via Pietro Giuria 1, I-10125 Turin, Italy}

\affiliation[b]{INFN, Turin,\\
Via Pietro Giuria 1, I-10125 Turin, Italy}

\emailAdd{elia.cellini@unito.it}
%\emailAdd{s.author@univ.country}

\abstract{Stochastic normalizing flows are a class of deep generative models that combine normalizing flows with Monte Carlo updates and can be used in lattice field theory to sample from Boltzmann distributions. In this proceeding, we outline the construction of these hybrid algorithms, pointing out that the theoretical background can be related to Jarzynski's equality, a non-equilibrium statistical mechanics theorem that has been successfully used to compute free energy in lattice field theory. We conclude with examples of applications to the two-dimensional $\phi^4$ field theory.}
\FullConference{%
The 39th International Symposium on Lattice Field Theory,\\
8th-13th August, 2022,\\
Rheinische Friedrich-Wilhelms-Universität Bonn, Bonn, Germany
}
%% \tableofcontents
\maketitle

\section{Introduction}
Following ref. \cite{Caselle:2022acb}, in this proceeding, we describe how to combine normalizing flows \cite{rezende2015variational} with Monte Carlo updates in a new class of generative models called stochastic normalizing flows \cite{wu2020stochastic} which can be easily described in a framework inspired by non-equilibrium statistical mechanics. Normalizing flows are a class of deep generative models used to efficiently evaluate approximations of statistical distributions by mapping them to suitable distributions. In lattice field theory, normalizing flows can sample uncorrelated configurations from Boltzmann distributions \cite{Albergo1} and a direct application is to use them to compute physically-interesting thermodynamic observables \cite{Nicoli2021}. Thus, this class of models provides a new, promising route to studying quantum field theories on lattice \cite{Albergo1,Nicoli2021,Albergo2,Albergo3,Hackett:2021idh,Abbott:2022zhs,DelDebbio,Pawlowski:2022rdn,Wynen:2020uzx,Finkenrath:2022ogg,deHaan:2021erb,Gerdes:2022eve}. Furthermore, stochastic normalizing flows are created by combining stochastic updates and normalizing flow layers and share the same theoretical background that underlies Monte Carlo simulations based on Jarzynski’s equality \cite{Jarzynski1997}, a method that found successful application in lattice field theory \cite{Caselle:2016wsw, Francesconi:2020fgi, CaselleSU3}.

%In the following, we summarize normalizing flows, out-of-equilibrium Monte Carlo simulations, and stochastic normalizing flows (Sec. (\ref{SNF})). We conclude with applications on lattice field theory (Sec. (\ref{testing})), comments and future applications (Sec. (\ref{conclusion})).

\section{Stochastic normalizing flows}\label{SNF}
In this section, before introducing stochastic normalizing flows \cite{wu2020stochastic}, we briefly summarize the building blocks of these hybrid algorithms: normalizing flows \cite{rezende2015variational} and out-of-equilibrium Monte Carlo simulations based on Jarzynski's equality \cite{Jarzynski1997}. A normalizing flow is a function $f: \mathbb{R}^d\to\mathbb{R}^d$, that expresses a sequence of invertible and differentiable transformations interpolating between a prior distribution $q_0(\phi_0), \; \phi_0 \in \mathbb{R}^d$ and the target distribution $p(\phi)=\exp(-S[\phi])/Z,\;\phi \in \mathbb{R}^d$. Normalizing flows can be implemented using neural networks by composing $N$ layers labeled by a natural number $0\leq n\leq N$. Using the change of variable formula it is possible to compute the density of the generated samples: $q(g(\phi_0))=q_0(\phi_0)|\det J_g(\phi_0)|^{-1}$, where $J_g$ is the Jacobian matrix associated with the map $g=f^{-1}$. Therefore, for practical implementation, $g$ must have a tractable determinant of the Jacobian. Normalizing flows can be trained so that the "learned" distribution $q$ well approximates the target $p$. The training is done by minimizing the Kullback-Leibler divergence: $D_{KL}(q||p)=\int d\phi [\ln q(\phi)-\ln p(\phi)]$ which is a measure of similarity between the $q$ and $p$. After training, the partition function of the target $p$ can be computed using a reweighting procedure, also called importance sampling in the machine learning field \cite{Nicoli2021}:
\begin{equation}\label{NF_Z}
    \frac{Z}{Z_0}=\int d\phi q(\phi)\tilde{w}(\phi)=\langle \Tilde{w}(\phi)\rangle_{\phi \sim q}
\end{equation}
where we introduce the weight:
\begin{equation}\label{flow_weight}
    \Tilde{w}(\phi_0)=\exp\bigl(-\bigl\{S[g(\phi_0)]-S_0[\phi_0]-Q_g\bigr\}\bigr)
\end{equation}
and $q_0(\phi_0)=\exp(-S_0[\phi_0])/Z_0$. The form of $Q_g=\sum_{n=0}^{N-1}\ln |\det J_n(\phi_n)|$ depends on the network architectures chosen to implement the layers $g_n$. Normalizing flow layers can be combined with Monte Carlo methods to obtain hybrids frameworks, following ref. \cite{Caselle:2022acb}, we compare normalizing flows with the theoretical background of Jarzynski's equality\cite{Jarzynski1997}: a non-equilibrium statistical mechanics theorem successfully applied for free energy calculations in Monte Carlo simulations of lattice gauge theories \cite{Caselle:2016wsw, Francesconi:2020fgi,CaselleSU3}. Jarzynski's equality states that equilibrium partition functions ratios can be calculated as an exponential average over non-equilibrium processes of the dimensionless work $w$ done on the system:
\begin{equation}\label{Jarzynski}
    \frac{Z_{\eta_{fin}}}{Z_{\eta_{in}}}=\langle \exp(-w(\phi_0,\phi_1,...,\phi_N))\rangle_f 
\end{equation}
where we introduced a protocol $\eta(t)$ that drives out of equilibrium the states $\eta_{in}=\eta(t_{in})$ to $\eta_{fin}=\eta(t_{fin})$ and can be a set of couplings appearing in the action $S$. The average of the equation (\ref{Jarzynski}) is taken over all possible trajectories connecting, in the phase space, $\eta_{in}$ to $\eta_{fin}$. Note that only the initial configurations must be in the thermodynamic equilibrium; the system is forced out of equilibrium and is never allowed to relax. These \textit{stochastic evolutions}\footnote{An equivalent algorithm is the \textit{annealed importance sampling} \cite{neal2001annealed}} can be implemented using Monte Carlo updates by discretizing the time $t$ and computing the work as:
\begin{equation}\label{eq:JE_work}
    w(\phi_0,\phi_1,...,\phi_N)  = S_{\eta_N}[\phi_N]-S_{\eta_0}[\phi_0]-Q_h(\phi_0,\phi_1,....,\phi_N) 
\end{equation}
where $\phi$ are the degrees of freedom of the system and $\eta_{n}=\eta(t_{n})$ and $Q_h$ is the heat exchanged with the environment during the processes. Given a discrete protocol $\eta$, it is possible to define a sequence of Boltzmann distributions, at the $n$-th step: $\pi_{\eta_{n}}[\phi]=\exp(-S_{\eta_n}[\phi])/Z_{\eta_n}$, the "prior" $\phi_0$ is sampled from the distribution $\pi_{\eta_{0}}=e^{-S_{\eta_{0}}[\phi_0]}/Z_{\eta_{0}}$ and the heat can be computed as: $Q_h(\phi_0,\phi_1,...,\phi_N)=\sum_{n=0}^{N-1}\{ S_{\eta_{n+1}}[\phi_{n+1}]-S_{\eta_{n+1}}[\phi_{n}] \}$. Note that eq. (\ref{Jarzynski}) doesn't depend on protocol $\eta(t)$. However, there is a limited number of trajectories in the Monte Carlo implementation; hence, the protocol choices impact the overall efficiency of the method. Finally, stochastic normalizing flows can be constructed interleaving normalizing flow layers with stochastic evolution updates: in this hybrid framework partition functions can be calculated using Jarzynki's equality (\ref{Jarzynski}) with generalized work $w(\phi_0,\phi_1,...,\phi_N)  = S_{\eta_N}[\phi_N]-S_{\eta_0}[\phi_0]-Q(\phi_0,\phi_1,...\phi_N)$ where the "heat" $Q$ is computed as the sum of $Q_g$ and $Q_h$. Moreover, generic observables $\mathcal{O}$ at $\eta=\eta_{fin}$ can be computed using Jarzynski’s equality:
\begin{equation}\label{eq:SNFobservable}
    \langle \mathcal{O}\rangle_{\eta=\eta_{fin}}=\frac{\langle \mathcal{O}(\phi_N) \exp(-w(\phi_0,\phi_1,...,\phi_N))\rangle_f}{\langle \exp(-w(\phi_0,\phi_1,...,\phi_N))\rangle_f}.
\end{equation}

%The reader may have noticed the remarkable similarity between normalizing flows and stochastic evolutions. Both provide maps across distributions. Thus if the configurations generated according to the prior distribution are uncorrelated, the generated samples present no correlations. Furthermore, neither method seeks to model the exact target distribution but rather to find an approximation that can be used as a highly expressive prior for the reweighing procedure or the Metropolis-Hastings algorithm. In the case of stochastic evolution, we also have a physics interpretation of the approximation, which is an out-of-equilibrium perturbation of the target. Moreover, the relations between  $w$ (\ref{eq:JE_work}) and  $-\ln\Tilde{w}$ (\ref{flow_weight}) are explicit and nothing prevent us to create that stochastic normalizing flows by concatenating normalizing flows layers and stochastic evolution updates. In this hybrid framework, the partition functions can be calculated using Jarzynky's equality \ref{Jarzynski} with generalized work $w(\phi_0,\phi_1,...,y_N)  = S[y_N]-S_{\eta_0}[\phi_0]-Q(\phi_0,\phi_1,...,y_N)$ where the heat $Q$ is computed as the sum of $Q_g$ and $Q_h$. 

%The original idea of combining deterministic with stochastic steps is by \cite{Vaikuntanathan_2011}. Nevertheless, deep learning has now given these hybrid methods a novel perspective \cite{wu2020stochastic, Caselle:2022acb, Matthews:2022sds}. 

\section{Application to scalar field theory}\label{testing}
In this section, we show some results in the two-dimensional $\phi^4$ interacting field theory, more extended studies can be found in \cite{Caselle:2022acb}. The theory is regularized on a lattice $\Lambda$ of size $L_t\times L_s$, with lattice spacing $a$ and periodic boundary conditions along both dimensions. We defined $N_t=L_t/a$ and $N_s=L_s/a$ as the number of sites in the temporal and spatial directions. The Euclidean action is:
\begin{equation}\label{action}
    S(\phi)=\sum_{x\in \Lambda}-2\kappa\sum_{\mu=0,1}\phi(x)\phi(x+\hat{\mu})+(1-2\lambda)\phi(x)^2+\lambda\phi(x)^4.
\end{equation}
For each algorithm: normalizing flows (NFs), stochastic evolutions (SEs) and stochastic normalizing flows (SNFs), we sample the prior from a normal distribution with $\mu=0$ and $\sigma=0.5$, thus we recover (\ref{action}) with $\kappa=0$ and $\lambda=0$. This choice simplifies the protocol needed for SEs and SNFs. To compare the performances, we use the effective sample sizes (ESS):
\begin{equation}
    ESS=\frac{\langle\Tilde{w}\rangle_{f}^2}{\langle\Tilde{w}^2\rangle_{f}}
\end{equation}
which is defined in the interval $[0,1]$ and is equal to $1$ when the generated distribution $q$ is equal to the target.
\begin{figure}
  \centering
  \includegraphics[scale=0.9,keepaspectratio=true]{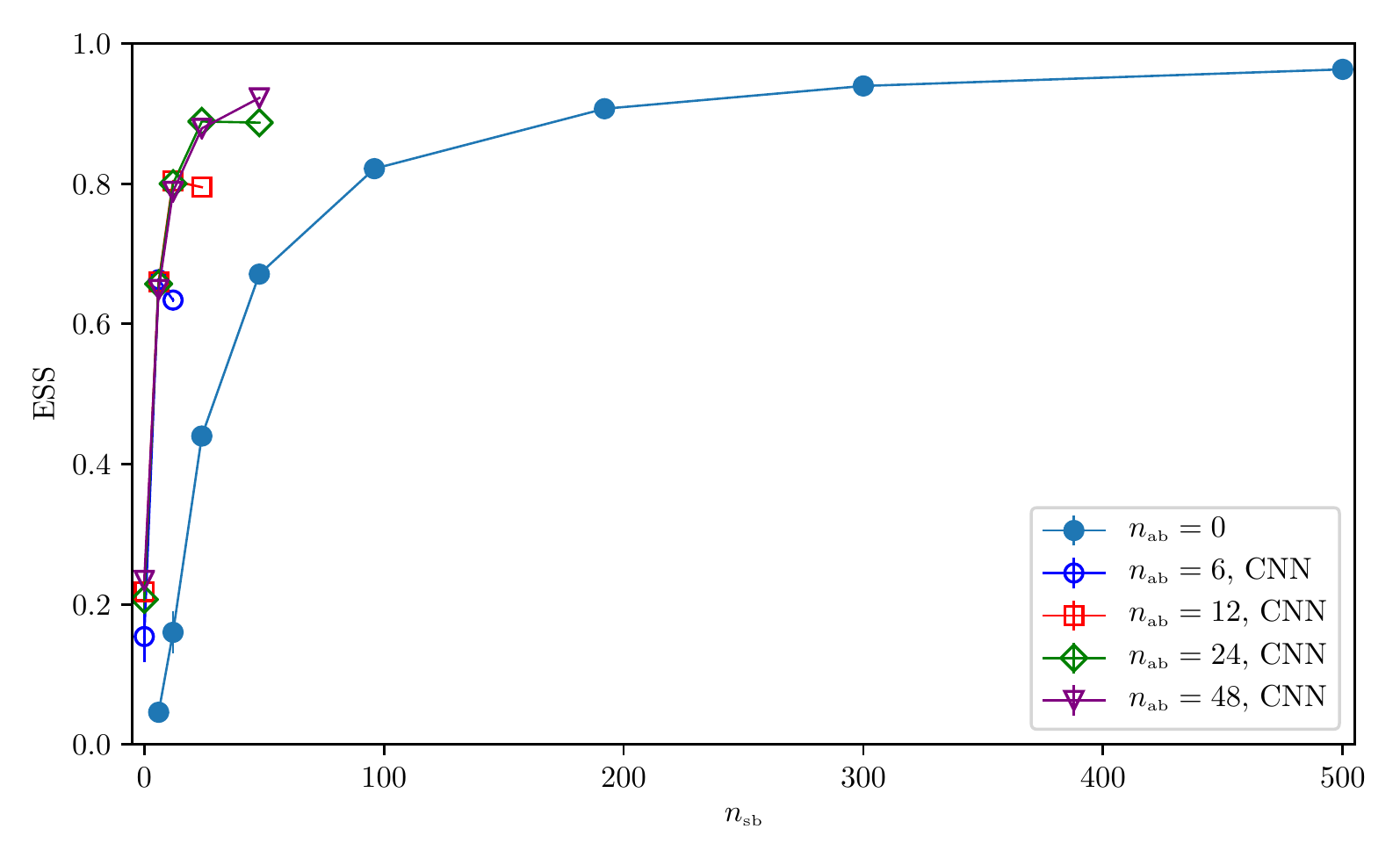}
  \caption{ESS as a function of $n_{sb}$ for different flows. $16\times 8$ lattices at $\kappa=0.2,\; \lambda=0.022$. The filled $n_{ab}=0$ point represents SEs, while the empty points represent SNFs. In the limit $n_{sb}=0$ and $n_{ab}\neq 0$ NFs are recovered.}
  \label{fig:arch}
\end{figure}

For normalizing flow layers, we implement the RealNVP \cite{dinh:2017} affine coupling layers and enforce $\mathbb{Z}_2$ equivariance as \cite{DelDebbio}. The networks used are minimal convolutional neural networks with just one layer, two feature maps, and $3\times 3$ kernel. Since each coupling layer transforms only half of the lattice sites (we use a "checkerboard" even-odd, partitioning), we define one affine block as the sum of two affine layers and introduce $n_{ab}$ as the number of deterministic, affine blocks. For SEs, we fix a linear protocol $\eta$ interpolating between the initial and final action parameters, the $n$-th "layer" is defined by the protocol parameters $\eta_n=\eta(t_n)$ that are used to update the configurations with the action $S_{\eta_{n}}$. We implemented local heatbath updates and defined the number of stochastic blocks as $n_{sb}$. SNFs are built by alternating affine blocks with stochastic updates. The models (NFs and SNFs) are trained by minimizing the loss function $-\langle \ln \Tilde{w}\rangle_{f}$ which is the Kullback-Leibler divergence minus $\ln Z/Z_0$. To update the model parameters, we use ADAM \cite{Kingma:2014vow} with $10^4$ steps, set the initial learning rate to $0.0005$, and use ReduceLROnPlateau scheduler with patience of $500$ steps. The software used is written using Pytorch \cite{Pythorc}. All the numerical experiments have been performed on an NVIDIA Volta V100 GPU with 16 GB.

\begin{figure}
  \centering
  \includegraphics[scale=0.9,keepaspectratio=true]{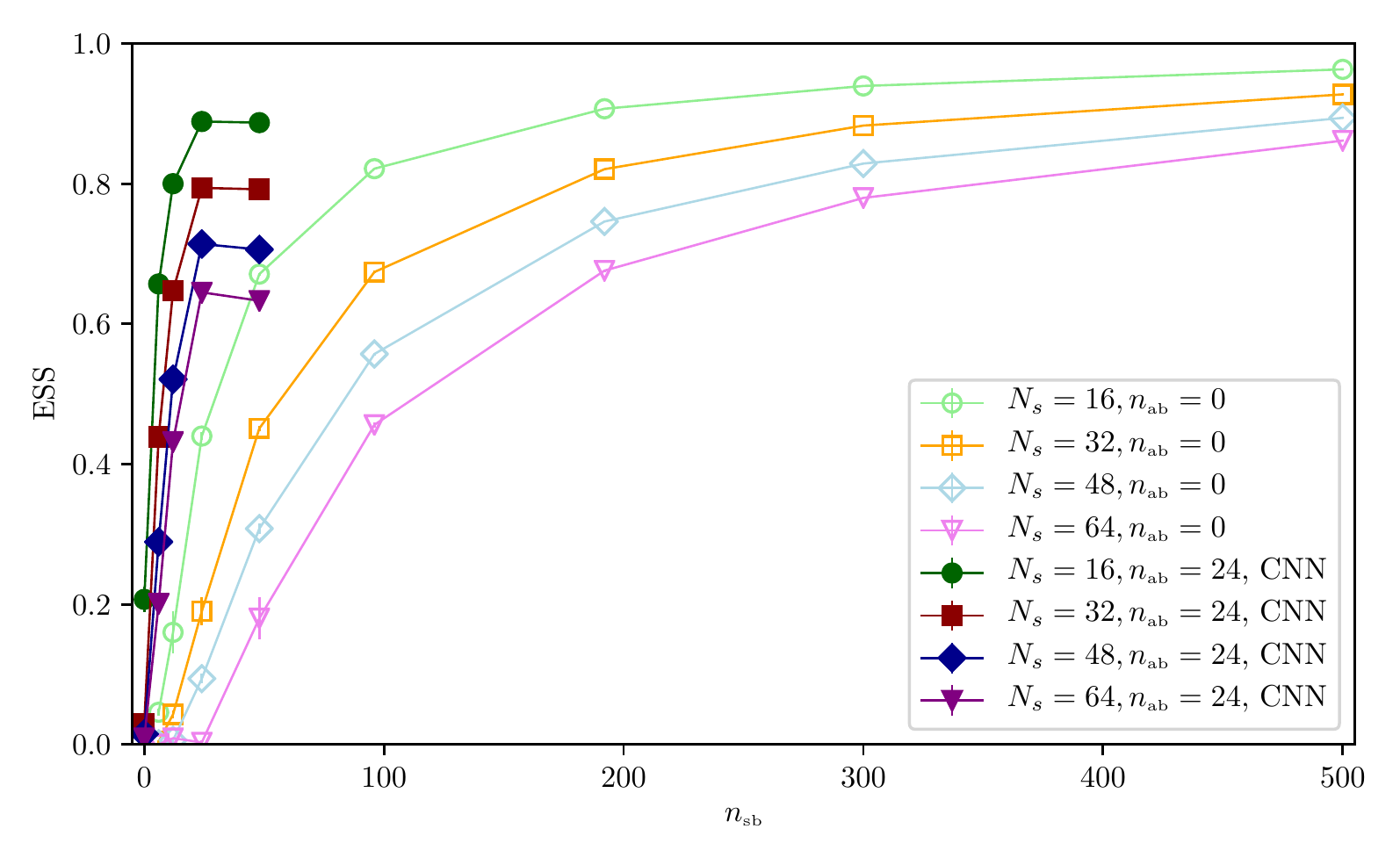}
  \caption{ESS as function of $n_{sb}$ and $N_s$ for SEs and SNFs with $n_{ab}=24$. Lattices of size $N_t=8$ at $\kappa=0.2,\; \lambda=0.022$.}
  \label{fig:volume}
\end{figure}

\begin{figure}
  \centering
  \includegraphics[scale=0.5,keepaspectratio=true]{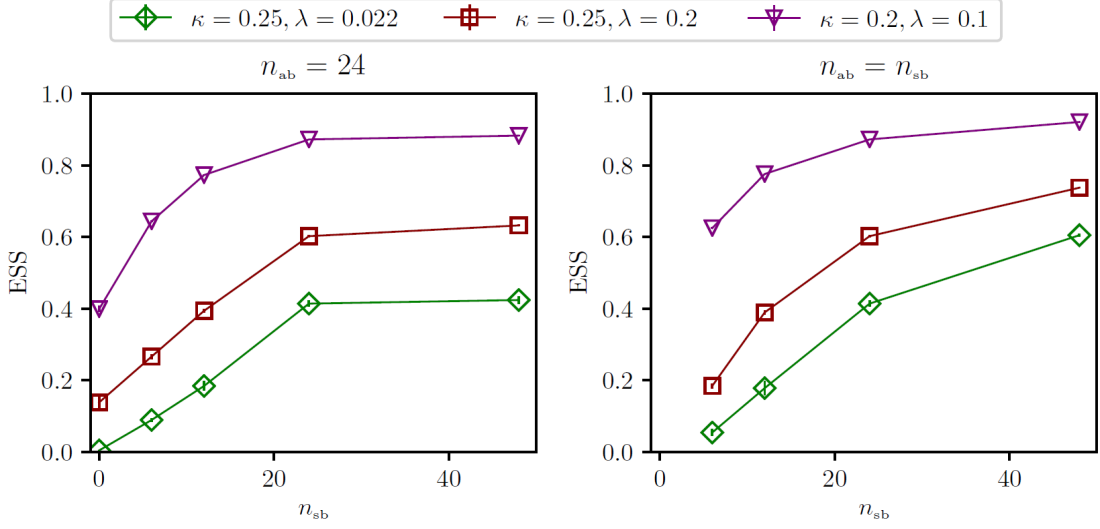}
  \caption{ESS as a function of $n_{sb}$  SNFs with $n_{sb}= 24$ (left panel) and $n_{sb}=n_{ab}$ (right panel). Lattices of size $16\times 8$ and different values of the target parameters. To the best of our knowledge, the green target parameters should lie very close to the critical point.}
  \label{fig:otherpar}
 \end{figure}

\begin{figure}
  \centering
  \includegraphics[scale=0.4,keepaspectratio=true]{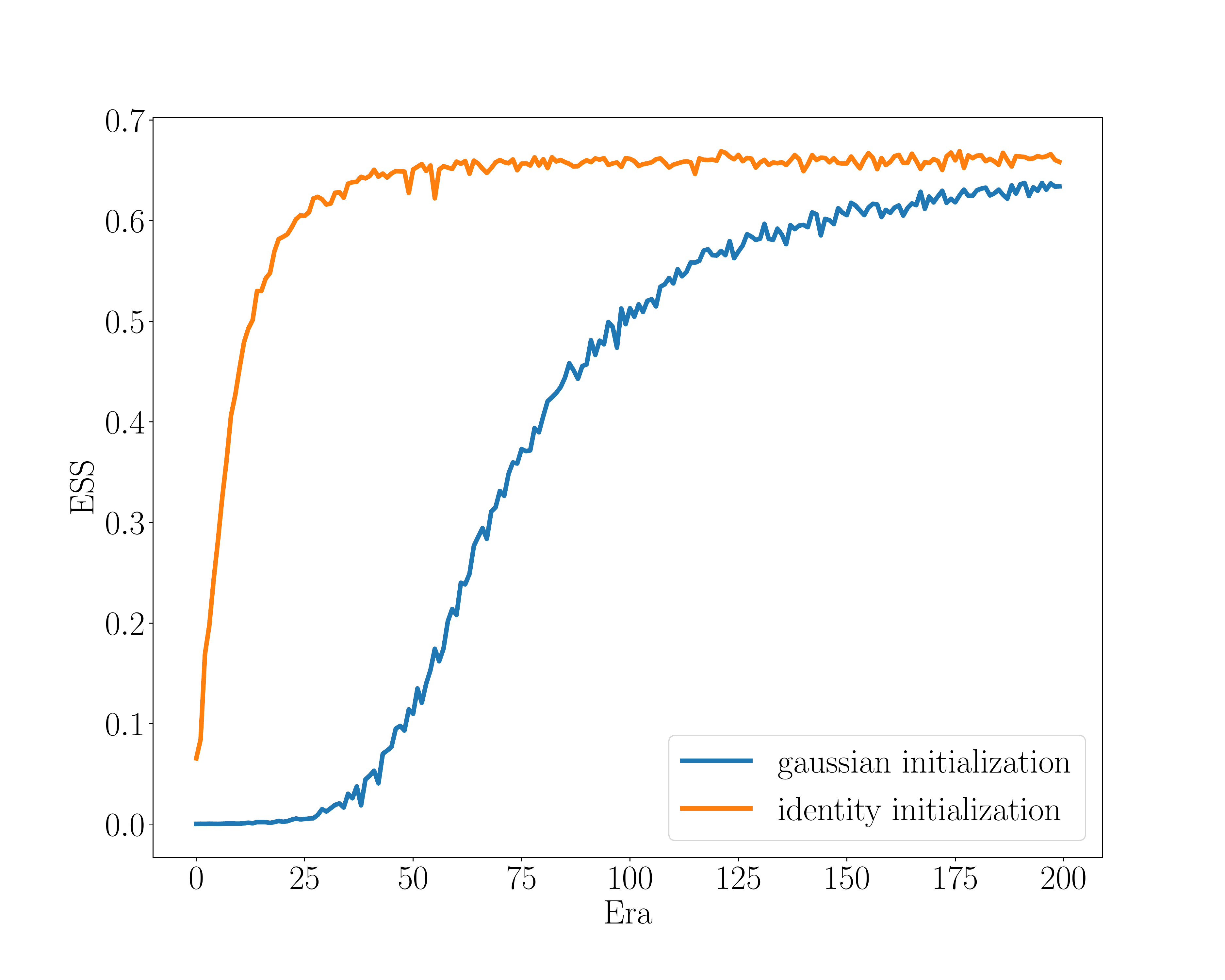}
  \caption{Learning curve of ESS for different SNFs initializations. One era corresponds to the average of 10 optimizer updates. $16\times 8$ lattices at $\kappa=0.2,\; \lambda=0.022$.}
  \label{fig:lc}
\end{figure}

In fig. (\ref{fig:arch}), we fixed the volume of the lattices to $16\times 8$ and set the couplings to $\kappa=0.2$ and $\lambda=0.022$, which lie in the symmetric phase of the theory. In fig. (\ref{fig:volume}) we fixed $n_{ab}=24$ and $N_t=8$ and we vary $n_{sb}$. Both fig. (\ref{fig:arch}) and (\ref{fig:volume}) show that SNFs can reach high ESS using a small number of $n_{sb}$; this is highly relevant because the number of stochastic updates is drastically reduced. We also observed a performance peak when $n_{sb}=n_{ab}$, which shows that the layer produced by one deterministic and one stochastic block is the most expressive ingredient of SNFs. 
 In fig. (\ref{fig:otherpar}), we target different action parameters finding the same trends as the other study, namely the saturation of the ESS when $n_{ab}>n_{sb}$. We remark that all the tests are performed in the symmetric phase; a different type of SNFs may be required to reproduce our results in the unbroken symmetry phase. In fig. (\ref{fig:lc}), we compare Gaussian and identity initializations for flow parameters; for all the measures in this proceeding, we initialized the parameters of flows with random Gaussian values. Furthermore, following ref. \cite{Matthews:2022sds}, we found that identity initialization for NF layers can speed training; this choice reduces the untrained SNFs to SEs, offering suitable starting points for training procedures.
\section{Conclusion}\label{conclusion}
In this proceeding, based on \cite{Caselle:2022acb}, we have discussed the relationship between normalizing flows and non-equilibrium Monte Carlo simulations related to Jarzynski's equality (stochastic evolutions): both, providing deterministic or stochastic maps, give an effective way to compute partition functions and sample target distributions. Moreover, stochastic normalizing flows can be constructed by combining normalizing flows and stochastic evolutions; in this hybrid framework, the non-equilibrium protocol helps deterministic blocks to find suitable paths, while normalizing flows provide highly expressive maps. Since Monte Carlo updates are ergodic, stochasticity ameliorates mode-collapsing of normalizing flows. Despite the potential of these novel algorithms, the accuracy of the measures depends on a wide range of largely undiscovered possibilities; further research is required, and out-of-equilibrium thermodynamics can be exploited as a theoretical tool to drive the investigation of novel algorithms.

Stochastic evolution has been successfully applied to high-precision lattice gauge theory studies \cite{Caselle:2016wsw, Francesconi:2020fgi, CaselleSU3}. Hence, natural developments of our contribution will examine the extension of these works using stochastic normalizing flows. We showed it is possible to use stochasticity to improve simple normalizing flows. Therefore, other improvements of this work include the study of the stochastic counterparts of state-of-art flow-based samplers like continuous normalizing flows \cite{deHaan:2021erb,Gerdes:2022eve} and the application to the lattice field theory of extensions of stochastic normalizing flows \cite{Matthews:2022sds}. 
\acknowledgments
The numerical simulations were run on machines of the Consorzio
Interuniversitario per il Calcolo Automatico dell'Italia Nord Orientale
(CINECA). We acknowledge support from the SFT Scientific Initiative of
INFN. This work was partially supported by the Simons Foundation grant 994300
(Simons Collaboration on Confinement and QCD Strings). Part of the numerical functions used in the present work are based
on ref.~\cite{Albergo:2021vyo}.

\bibliographystyle{JHEP}
\bibliography{biblio.bib}

\end{document}